\documentclass[11pt,reqno]{article}
\usepackage{geometry,verbatim}
\usepackage{caption}
\usepackage{subcaption}
\usepackage[english]{babel}
\usepackage{natbib}
\usepackage{amsmath,amsthm,amssymb,mathtools}
\usepackage{ulem,xpatch}
\xpatchcmd{\sout}
  {\bgroup}
  {\bgroup}
  {}{}

\usepackage{dsfont}
\usepackage{url}
\usepackage{lscape}
\usepackage{color}
\usepackage{setspace}
\usepackage{booktabs}
\usepackage[flushleft]{threeparttable}
\usepackage{pdfpages}
\usepackage{booktabs}
\usepackage{rotating,adjustbox}
\usepackage{multicol}
\usepackage{multirow}
\usepackage{array}

\usepackage[utf8]{inputenc}
\usepackage[T1]{fontenc}
\usepackage{enumitem}
\usepackage{hyperref}
\hypersetup{
    colorlinks=true,
    linkcolor=blue,
    urlcolor=blue,
}

\newcolumntype{L}[1]{>{\raggedright\let\newline\\\arraybackslash\hspace{0pt}}m{#1}}
\newcolumntype{C}[1]{>{\centering\let\newline\\\arraybackslash\hspace{0pt}}m{#1}}
\newcolumntype{R}[1]{>{\raggedleft\let\newline\\\arraybackslash\hspace{0pt}}m{#1}}
\usepackage{amssymb}
\usepackage{pdflscape}
\usepackage{hyperref}
\hypersetup{colorlinks=true, linkcolor=red,          
    citecolor=blue,        
    filecolor=magenta,      
    urlcolor=cyan           
}
\usepackage{tikz}

\usepackage{tikz-qtree}
\usetikzlibrary{arrows,shapes,positioning,shadows,trees,calc}

\tikzset{edge from parent/.style=
            {thick, draw, edge from parent fork right},
         every tree node/.style=
            {draw,minimum width=1in,text width=1.5in,align=center}}
\theoremstyle{plain}

\newtheorem{corollary}{Corollary}
\newtheorem{lemma}{Lemma}
\newtheorem{proposition}{Proposition}
\theoremstyle{definition}
\newtheorem{definition}{Definition}
\newtheorem{remark}{\textbf{Remark}}
\theoremstyle{remark}
\newtheorem{example}{\textbf{Example}}
\DeclareGraphicsExtensions{.eps,.ps,.eps.gz,.ps.gz,.eps.Z}

\usepackage{datetime}
\newdateformat{monthyeardate}{%
  \monthname[\THEMONTH], \THEYEAR}

\usepackage[english]{babel}
\usepackage{hyperref}

\addto\extrasenglish{%
}
\addto\extrasenglish{%
}
\usepackage{dashbox}
\newcommand\dboxed[1]{\dbox{\ensuremath{#1}}}
\title{Time-constrained Dynamic Mechanisms for College Admissions\thanks{We thank In\'{a}cio B\'{o}, Estelle Cantillon, Caterina Calsamiglia, Yan Chen, Joana Pais, the Associate Editor and anonymous referees for helpful comments. We are also indebted to participants at Department of Economics (dECON) - Uruguay, University of Gothenburg, UECE Lisbon Meetings, AMES Hong Kong, and the 10th Conference on Economic Design.
This project is supported by the Young Scientists Fund of the National Natural Science Foundation of China (Grant no. 72504207), Fundamental Research Funds for the Central Universities (Tongji University - Grant no. 22120260361),  the Belgian National Science Foundation (FNRS), European Research Council Grant n.339950, from ANII, FCE\_1\_2023\_1\_176072, and the Anniversary Foundation of the School of Business, Economics and Law, University of Gothenburg.}
}
\date{}
\author{
Li Chen\thanks
{Shanghai International College of Intellectual Property, Tongji University, China; Department of Economics, University of Gothenburg, Sweden. \url{chen_li@tongji.edu.cn}},
Juan S. Pereyra\thanks{Department of Economics, Universidad de Montevideo, Uruguay. \url{jspereyra@um.edu.uy}},
Min Zhu\thanks{Corresponding author. Department of Economics, Beijing Normal University, China. \url{zhu@bnu.edu.cn}}
}

\begin{document}

\maketitle

\begin{center}

\monthyeardate\today

\end{center}

\begin{abstract}
Recent literature shows that dynamic matching mechanisms may outperform their static counterparts. The college admissions procedure used in Inner Mongolia motivates an underexplored design dimension of such mechanisms: the time constraint that students face. We study the time-constrained dynamic mechanism (TCDM) theoretically and show that, under straightforward behavior, it can generate unstable outcomes and, in some cases, outcomes that are Pareto dominated by those produced by the student-proposing deferred acceptance mechanism (DA). We also compare students’ assignment probabilities under TCDM and DA. Under this behavior, TCDM gives each student a weakly higher probability of receiving her first choice. Under universal acceptability and sufficient aggregate capacity, it also gives each student a weakly higher probability of remaining unassigned and a weakly lower probability of receiving a non-first-choice college.
\end{abstract}


\clearpage

\section{Introduction}\label{intro}

Research in market design has successfully helped to design or redesign many real-world applications over the past decades, including school choice and college admissions \citep{roth1982economics,pathak2017really}. 
Under the standard approach, students submit a rank-ordered list of preferences, and the mechanism combines these reports with colleges' priorities and capacities to produce assignments. The mechanism is direct, as students submit their preferences directly to the mechanism, and it is static, as the submitted list is input to the mechanism and cannot be revised. 
Strategyproof mechanisms such as the student-proposing deferred-acceptance (DA) mechanism, introduced by \citet{gale62}, are endorsed by the literature because students can do no better than submitting preferences truthfully.  
Nevertheless, experimental and field evidence shows that strategyproof mechanisms may fail to convince all students to submit preferences truthfully in the lab and in the field. Such behavior may arise from, for instance, uncertainty \citep{pais2008school,hassidim2017mechanism, rees2018suboptimal,chen2019self}, the complexity of the choice problem \citep{shorrer2021dominated}, or richer preferences, including loss aversion and rank-dependent utility \citep{dreyfuss2022expectations,meisner2023loss,meisner2023report}. While some deviations from truthful reporting are consistent with equilibrium strategies, others could harm students and prevent the mechanisms from achieving desirable outcomes.

One explanation for such behavior is that students may not fully understand the mechanism. Providing students with information and opportunities to revise their choices may therefore help them make better-informed decisions. Recent studies, mostly experimental, show that dynamic mechanisms---in which tentative assignments are updated sequentially, and students can revise their choices after receiving information or feedback about their current allocations---may outperform their static counterparts \citep{stephenson2016continuous,ding2017matching,klijn2019static,bo2020iterative,gong2016dynamic}. 
While this evidence often comes from small markets, many markets in practice are considerably larger, and the sequential process could be extremely time-consuming. As a result, clearinghouses often impose a deadline by which students must finalize their submissions. The time available to students varies substantially across systems, ranging from hours to months (see Table \ref{tab:dynamic}). 

\begin{table}[htp]
    \caption{Dynamic mechanisms in practice}\label{tab:dynamic}
    \centering
    \footnotesize 
    \begin{tabular}{L{2cm}llL{3cm}L{3cm}}
\toprule 
     & Education system & Application window & Information & Source \\ \midrule
 Brazil &   University admissions  & 5 days&  University cutoffs & \cite{BoHa2016} \\  \hline
 Germany    &   University admissions   & 31 days & Applicant’s rank at each university to which she applies     &     \cite{grenet2022preference}   \\  \hline
 Inner Mongolia, China   &    University admissions  & 11 hours &  University and program cutoffs; applicant’s rank at the applied university and first-choice program &       \cite{gong2016dynamic} and authors’ data \\ \hline
 Wake County, US     &     School choice    &   2 weeks & Number of first-choice applicants     &    \cite{dur2015identifying}     \\  
     \bottomrule
\end{tabular}

\end{table}

Time constraints may, however, prevent dynamic mechanisms from achieving desirable outcomes if students do not have enough time to revise their choices. An example is the dynamic mechanism used for college admissions in Inner Mongolia, China, where the  procedure operates within a single day and students have a few hours to submit and finalize their choices. The initial application consists of one  university, which can be  revised as information about tentative assignments becomes available. Although the number of revisions is not capped, the mechanism imposes strict time constraints: students must finalize their choices by a predetermined deadline.\footnote{The Inner Mongolian system uses a staggered closing rule, dividing students into batches based on their scores. Higher-scoring students in an earlier batch finalize their choices first. Lower-scoring students from subsequent batches are given an additional hour to revise.} 
Table \ref{tab:outcome} presents application behavior and assignment outcomes using hourly assignment data published by the Inner Mongolia clearinghouse in 2018. The data reveal that 47.4\% (10,003) of STEM-track and  54.2\% (3,303) of humanities-track applicants changed their applications in the final round. Yet, 
5.7\% (1,208) of the STEM track and 9.5\% (577) of the humanities track were ultimately unassigned despite revising  in the final hour. In addition, 5.5\% (1,161) from the STEM track  and 5.3\% (325) from the humanities track were ultimately unassigned even though their exam scores were at or above the final cutoff of a university to which 
they had previously applied. This pattern provides suggestive evidence that a tight deadline may generate unstable outcomes---that is, outcomes in which some student prefers a college to the current assignment and that college either has a vacant seat or admits a lower-priority student.

Motivated by this evidence, we introduce a model of centralized college admissions in which stability is desirable. Students have strict preferences over colleges, and colleges share a common strict priority order determined, for instance, by exam scores. With this common-priority structure, DA selects the unique stable matching, which is also Pareto efficient.  

We study a mechanism where students initially submit a single application, generating a tentative matching. In subsequent rounds, students may revise their applications in response to updated information about tentative assignments and college cutoffs. In particular, this information allows each student to identify the colleges that remain attainable to her, that is, colleges whose current cutoffs do not exceed her score. Applications are submitted simultaneously within each round. When the deadline is reached, the final tentative matching is implemented. Because the process may end before students have enough time to complete their revisions, we refer to it as the time-constrained dynamic mechanism (TCDM).

\begin{table}[tp]
  \caption{Application behavior and assignment outcomes (2018)}\label{tab:outcome}
  \centering
  \small
\begin{threeparttable}
\begin{tabular}{p{7.5cm}p{3cm}p{3cm}}
\toprule
  & STEM &   Humanities \\
\cmidrule{1-3}
Number of students who changed in the final round & 10,003 & 3,303 \\
Number of students unassigned despite final-round change &  1,208  &  577 \\
Number of unassigned students with a score $\geq$ the final cutoff of a previously applied university & 1,161  & 325 \\
\cmidrule{1-3}
Total number of students & 21,107 & 6,089 \\
\bottomrule
\end{tabular}
\begin{tablenotes}[flushleft]
    \item \small Notes: The table reports application and assignment outcomes in the 2018 Inner Mongolian college admissions, for both STEM- and humanities-track applicants to top-tier institutions. The two tracks are processed separately using the same mechanism. Further
details on the data are provided in \autoref{data}.
\end{tablenotes}
\end{threeparttable}
\end{table}

To facilitate the comparison between the two mechanisms, we focus on \textit{straightforward strategies}: each student initially applies to the most-preferred option and, whenever rejected, applies to the most-preferred attainable option. Tentatively accepted students retain their applications.

Under straightforward behavior, TCDM can be interpreted as a compressed and time-truncated implementation of DA in which disclosed assignments and cutoffs let students skip colleges that would reject them, while the deadline may stop the process early. Unlike the truthful reporting that is weakly dominant under DA, straightforward behavior is not generally a dominant strategy under TCDM, and it need not constitute an equilibrium when the number of rounds is insufficient for the application-and-rejection process to reach the stable matching. We show, however, that straightforward behavior reaches the unique stable matching within $|I|$ rounds and that, when $T\geq |I|$, the profile of straightforward strategies is a Nash equilibrium.

Turning to the comparison with DA, our first result shows that if a student prefers her assignment under TCDM to her assignment under DA, then some higher-priority student remains unassigned under TCDM. The intuition is as follows. If a student prefers her assigned college under TCDM to the one assigned under DA, it must be the case that under DA, a higher-priority student occupies her place at that college. Consequently, either this higher-priority student is unassigned under TCDM, or she is assigned to a preferred college. Because the number of students is finite, this rejection-chain argument eventually identifies a higher-priority student who is unassigned when the TCDM stops. Consequently, a binding deadline may generate an unstable outcome. This does not by itself imply Pareto inefficiency: in general, the TCDM and DA outcomes may be Pareto incomparable, since some students may benefit from truncation while others are harmed. Nevertheless, TCDM may be Pareto inefficient, and DA can Pareto dominate it. Conversely, TCDM cannot Pareto dominate DA, because the DA outcome is Pareto efficient under the common priority order.

Second, we compare assignment probabilities under TCDM and DA from an ex-ante perspective. Under any distribution over preference profiles, TCDM gives each student a weakly higher probability of receiving her first choice. Under universal acceptability and sufficient aggregate capacity, it also gives each student a weakly higher probability of remaining unassigned and a weakly lower probability of receiving a non-first-choice college. These conclusions follow from profile-by-profile comparisons and therefore hold under any distribution over preference profiles. 
Finally, we show that, for each student, the probability of receiving her first choice under the $T$-round TCDM is weakly higher than under the $(T+1)$-round TCDM. In contrast, the expected number of students assigned to a non-first-choice college is weakly lower under the $T$-round TCDM than under the $(T+1)$-round TCDM.

We also consider an extension in which colleges impose different objective eligibility requirements while preserving the common priority order among eligible students. The main structural results are robust to this extension: under straightforward behavior, TCDM still converges to the unique stable matching within $|I|$ rounds, the profile of straightforward strategies remains a Nash equilibrium when $T\geq |I|$, and the rejection-chain characterization of students who benefit from truncation continues to hold.  Some ex-ante results are also robust: TCDM continues to weakly increase each student's probability of receiving her first choice relative to DA, and the comparative statics with respect to the number of rounds remain unchanged. 

\paragraph{Related Literature.}
We contribute to the literature by studying how a binding deadline affects an iterative implementation of the deferred acceptance mechanism. Starting with the seminal analysis of college admissions by \citet{gale62}, DA has become a workhorse in school choice \citep{abdulkadiroglu2003school} and college admissions \citep{balinski1999tale}.  A more recent literature compares the dynamic variants with the static DA, finding experimentally that dynamic mechanisms can produce more stable outcomes, lead to higher efficiency, and induce higher rates of truthful reporting \citep{bo2020iterative, gong2016dynamic,klijn2019static, stephenson2016continuous,dur2021sequential}. 

The experiment of \citet{bo2020iterative} is particularly related to our focus on the time available for iterative adjustment. Their experiment operates for a finite duration, and their results show that stable outcomes become more frequent as the iterative process unfolds. This evidence suggests that the amount of time available for revisions can materially affect outcomes. Our analysis complements their experimental findings by modeling the deadline explicitly and deriving profile-by-profile theoretical comparisons between a process stopped after $T$ rounds and the DA benchmark. In particular, we analyze the effects of truncation on stability, unassignment, assignment ranks, and the incentives to follow straightforward behavior.

\citet{BoHa2016} distinguish between the iterative mechanisms used in practice in Brazil and Inner Mongolia and the Iterative Deferred Acceptance Mechanism (IDAM) that they propose. Our TCDM is a stylized finite-round version of the former type of mechanism rather than of IDAM itself. Under TCDM, every student may revise her application in every round, including while tentatively assigned, and a new application replaces her preceding application. In the one-choice version of IDAM, by contrast, a student is asked to choose again only after her current choice is no longer held, her choice is restricted to an endogenous menu of available options, and offers are processed cumulatively. Thus, the two mechanisms have different strategy spaces and off-path incentive properties.

Under straightforward behavior, however, TCDM and the exam-based, one-choice version of IDAM generate closely related application paths: tentatively assigned students retain their applications, rejected students move to their most-preferred attainable options, and options that would reject them immediately are skipped. Accordingly, \autoref{NTC} in our paper is a common-priority counterpart of the convergence result in \citet{BoHa2016} (Proposition 1), with the additional explicit bound of $|I|$ rounds. The equilibrium results are less directly comparable. \citet{BoHa2016} (Theorem~1) establish that straightforward
type-strategies form an ex-post equilibrium in the unbounded IDAM. Our \autoref{NTC2} establishes a complete-information Nash equilibrium for TCDM when $T\geq |I|$, despite the fact that TCDM permits tentatively assigned students to revise their applications. Our main analysis then concerns the case in which the deadline binds, where straightforward behavior may fail to be an equilibrium and the process may stop before reaching the stable matching.

In another closely related paper, \citet{gong2016dynamic} model revision opportunities using individual Poisson clocks with parameter $\lambda$. In particular, each student has an individual Poisson clock and can change her application every time the clock rings. Thus, if the mechanism runs up to time $T$, the expected number of revision opportunities for each student is $\lambda T$. \citet{gong2016dynamic} assume that $\lambda$ is sufficiently large, so that students are effectively unconstrained in the number of revisions they can make.

There is a growing literature that evaluates sequential implementation and mechanism-generated information using field data. 
\citet{dur2015identifying} also study a dynamic mechanism where students can revise their reports before the final assignment. During the coordination phase, students do not observe tentative assignments; instead, they observe the number of students who rank each school first. The Boston mechanism (also known as the immediate acceptance mechanism) is then used to determine the final assignment. They find that sophisticated students are assigned to higher--performing schools. \citet{grenet2022preference} study a multi-offer dynamic mechanism in the German university admissions system and find that the combination of a sequential mechanism and a direct mechanism helps students discover their preferences. \citet{luflade2018value} finds that in Tunisian university admissions, offering information about vacancies to students under DA with a restricted number of choices can generate an outcome similar to that of unrestricted DA. \citet{kang2023dynamic} show empirically that replacing immediate acceptance with the dynamic mechanism in Inner Mongolia reduces justified envy but still performs worse than the static parallel mechanism. Our analysis provides a theoretical explanation of why justified envy may persist in the dynamic setting. 
Further, using a regression discontinuity design around the batching cutoff, \citet{kang2025information} find that students with lower priority within a batch tend to apply to safer, higher-capacity universities, leading them to enroll at universities with lower-performing peers.\footnote{ \citet{kovavc2025college} document a related pattern in the repeated DA procedure used in Croatian college admissions where applicants who receive a negative signal about their admission chances tend to switch to a more cautious choice.}  

When students do not have enough time to revise their applications, TCDM also resembles DA with constrained rank-order lists \citep{calsamiglia2010constrained,haeringer2009constrained}. Conditional on the final-round application profile, the last round of TCDM is equivalent to a constrained-DA mechanism in which each student lists only one option. The overall outcomes of the two mechanisms generally differ, however, because TCDM allows students to revise their applications over time and to skip colleges whose current cutoffs exceed their scores.

Our paper is also connected to the literature on decentralized sequential matching. For example, \citet{haeringer2011decentralized} study a sequential matching game in which firms make offers and workers decide whether to accept them. Their analysis focuses on strategic behavior in decentralized markets with commitment; in their benchmark environment, the unique subgame-perfect equilibrium outcome is the worker-optimal stable matching. Our environment differs in three respects: applications are made by students, colleges tentatively hold applicants according to priorities, and the process may stop before convergence. Thus, whereas they study equilibrium outcomes in a sequential offer game, we study the allocation consequences of imposing an application deadline on a dynamic implementation of DA under straightforward behavior.

Finally, our probability comparisons are also related to \citet{roth1999truncation}, who study truncation strategies in matching markets under symmetric incomplete information. They show that, under the firm-proposing stable mechanism, non-truncation strategies are stochastically dominated by truncations of true preferences, whereas different truncation levels are not generally comparable. A shorter list may increase the probability of obtaining a highly ranked match, but it may also increase the probability of remaining unassigned. An application deadline in TCDM generates an analogous trade-off endogenously. Students do not literally submit truncated lists, but the finite number of rounds limits how far down their preference rankings they can move. Thus, relative to DA, TCDM may increase the probability of receiving their first choice while also increasing the probability of remaining unassigned and reducing the probability of reaching lower-ranked options.

The remainder of the paper is organized as follows. We introduce the model setup and related mechanisms in \autoref{sec: model}. 
\autoref{with_const} presents the main theoretical results. We conclude in \autoref{sec: conclusion}. The appendices contain the proofs and additional material.

\section{The model}\label{sec: model}

There is a finite set of students $I$ to be assigned to a finite set of colleges $C$. Each student $i\in I$ has a strict preference order $\succ_i$ over $C\cup\{\emptyset\}$, where $\emptyset$ denotes the outside option. We say that a college $c$ is \textbf{acceptable} to student $i$ if $c\succ_i\emptyset$, and that there is \textbf{universal acceptability}
if every college is acceptable to every student, that is,
\[
c\succ_i\emptyset
\qquad\text{for every } i\in I \text{ and } c\in C.
\]
Let $\succeq_i$ denote the associated weak preference relation, and let
\[
\succ=(\succ_i)_{i\in I}
\]
denote a preference profile. 

Each college $c\in C$ has capacity $q_c\in\mathbb{N}_+$. We say that there is \textbf{sufficient aggregate capacity} if
\[
\sum_{c\in C} q_c \geq |I|.
\]

Each student $i\in I$ obtains a score $s_i>0$ from a centralized exam. We assume that scores are distinct, so they induce a common strict priority order
$\triangleright$ across colleges:
\[
i\triangleright j
\quad\Longleftrightarrow\quad
s_i>s_j.
\]
We extend our analysis to college-specific eligibility in
\autoref{subsec:eligibility}, allowing colleges to impose different objective eligibility criteria while maintaining the common exam-based priority order among eligible students.


A \textbf{matching} is a function
\(
\mu:I\rightarrow C\cup\{\emptyset\}
\)
such that
\[
|\mu^{-1}(c)|\leq q_c
\qquad\text{for every }c\in C,
\]
where
\[
\mu^{-1}(c)\equiv\{i\in I:\mu(i)=c\}
\]
is the set of students assigned to college $c$.

A student $i$ blocks a matching $\mu$ if $\emptyset \succ_i \mu(i)$.  
A student--college pair $(i,c)$ blocks a matching $\mu$ if
$c\succ_i\mu(i)$ and either
\(
|\mu^{-1}(c)|<q_c,
\)
or there exists $j\in\mu^{-1}(c)$ such that $i\triangleright j$.
A matching is \textbf{stable} if it is not blocked by a student or by a student-college pair. A matching $\mu$ is \textbf{Pareto efficient} if there is no matching $\mu'$ such that
\[
\mu'(i)\succeq_i\mu(i)
\quad\text{for every }i\in I,
\]
and $\mu'(j)\succ_j\mu(j)$ for a student $j\in I$.

Given a matching $\mu$, we define the \textbf{cutoff of college $c$} as
\[
f_c(\mu)=
\begin{cases}
\min\{s_i:i\in\mu^{-1}(c)\},
    & \text{if }|\mu^{-1}(c)|=q_c,\\
0,   & \text{if }|\mu^{-1}(c)|<q_c.
\end{cases}
\]


\subsection{Mechanisms}
We next describe two allocation mechanisms. We first present the student-proposing DA, which serves as our benchmark. Then, we turn to the TCDM. 

\paragraph{The student-proposing deferred acceptance mechanism (DA).}
DA is a direct mechanism where students submit a rank-ordered preference list once and the mechanism finds a matching without further action by students. It proceeds as follows: 

\begin{itemize}
    \item Step 1: For each student with at least one acceptable college, her most-preferred acceptable college is considered. Each college tentatively holds its highest-priority applicants up to capacity and rejects all other applicants.

    \item Step $d>1$: For each student rejected in the preceding step, her next most-preferred acceptable college not yet considered is considered. Each college considers its newly arriving applicants together with the students it tentatively held in the preceding step. It tentatively holds the highest-priority students up to capacity and rejects all others.

    \item The process ends when no further rejection occurs. Tentative acceptances then become final. A student who exhausts all colleges ranked above the outside option $\emptyset$ remains unassigned.
\end{itemize}

DA therefore runs the application process on behalf of students. It induces a static preference-revelation game in which truthful reporting is a weakly dominant strategy for students \citep{dubins1981machiavelli,roth1982economics}. To allow for comparison, we will assume in the rest of the paper that students submit preferences truthfully under DA.  

Under the common priority order, there is a unique stable
matching, which is Pareto efficient and can be found by DA. In addition, this matching coincides with the outcome of college-proposing DA and the outcome of serial dictatorship following the common priority order. Although these procedures select the same matching under the common priority order, we take DA as the benchmark.  As will become clear, under the application behavior we focus on, the TCDM operates as a compressed and time-truncated implementation of DA. 

\paragraph{The time-constrained dynamic mechanism (TCDM).}
TCDM is an indirect mechanism with $T\geq 1$ rounds. Instead of submitting a full preference list, each student chooses one college or the outside option in each round, and the application profile of the terminal round determines the final matching. Each round's applications generate a tentative matching, which the mechanism reveals to students together with the college cutoffs of that round. 

For each round $t$, let $\mu^{t}$ denote the tentative matching at the end of round $t$, and write $f_c^{t}\equiv f_c(\mu^{t})$ for the cutoff after round $t$.

The mechanism operates as follows: 

\begin{itemize}
    \item Round 1:  Each student applies to one college or chooses $\emptyset$. Based on the current application profile, each college tentatively holds its highest-priority applicants up to capacity and rejects all other applicants. The mechanism then reveals the tentative assignments and the current college cutoffs.

    \item Round $1<t\leq T$: Each student may retain or revise her preceding application. Based on the resulting current application profile, each college tentatively holds its highest-priority applicants up to capacity and rejects all others. The mechanism then updates and reveals the tentative assignments and college cutoffs.

    \item After Round $T$, the tentative matching becomes final.
\end{itemize}

The extensive-form game induced by TCDM has imperfect information because students do not observe other students' actions within the current round. In a given round $t$, a student observes the applications submitted in preceding rounds and the college cutoffs after the most recently completed round, but not the applications submitted by other students in the current round (a formal description of the extensive-form game induced by TCDM is presented in \autoref{app: extensive}).\footnote{Following the previous literature, we model students as observing the application history generated in earlier rounds. This information structure is stronger than necessary for our main results: under the common-score priority structure, the vector of cutoffs from the preceding round, together with a student's own score and tentative assignment status, is sufficient to implement straightforward behavior. Thus, the convergence and outcome-comparison results below would be unchanged if students observed only colleges' cutoffs.}  
In contrast to DA where truthful reporting is a weakly dominant strategy, students need not have dominant strategies, as in the sequential mechanisms studied in the literature \citep{klijn2019static,BoHa2016}. 
The comparison between TCDM and DA therefore requires fixing application behavior under TCDM, which we turn to next. 

\subsection{Application behavior}
We consider the following straightforward strategy, first introduced by \citet{BoHa2016}, which is a natural extension of truthful reporting under DA to our setting. Below, we provide an intuitive definition of this strategy to facilitate a clearer understanding (see \autoref{app: extensive} for a formal definition). 

A college is \textbf{attainable} for student $i$ in Round $t>1$ if $s_i \geq f_c^{t-1}$. That is, a college $c$ is attainable for student $i$ if $c$ has vacancies
after Round $t-1$, student $i$ is tentatively assigned to $c$, or student $i$ has higher priority than some student tentatively assigned to $c$. Since scores are distinct, the inequality is strict whenever student $i$ is not tentatively assigned to $c$. The outside option $\emptyset$ is attainable in every round, and every option is attainable in Round 1. 
\smallskip 

\begin{definition}(Straightforward strategy)
\label{def:straightforward}
In Round 1, student $i$ applies to her most-preferred option in $C\cup\{\emptyset\}$. In each Round $t>1$, student $i$ retains her current application if tentatively assigned. If unassigned after Round $t-1$, the student applies to her most-preferred attainable option. 
\end{definition}
\smallskip 
   
TCDM itself places no restriction on revisions. In every round, 
every student may revise her current application, even when she is tentatively accepted. This is a feature of the mechanism and therefore of the strategy space. Straightforward strategy, however, is more restrictive: a student retains her current application until she is rejected. Along the straightforward path, the mechanism therefore implements the application-and-rejection process of  DA,  though the limited number of rounds may stop completion of the process. Meanwhile, the revealed information speeds the process up, as a rejected student  can skip colleges that are unattainable.  

\begin{lemma}\label{lemma:monotone}
Suppose students follow straightforward strategies. Then, for every college $c$, cutoffs weakly increase over rounds: $f_c^{1}\leq f_c^{2}\leq\cdots\leq f_c^{T}$. 
\end{lemma}

To see why, first suppose that college $c$ is under capacity. In this case, its cutoff is zero. Once college $c$ is filled to capacity, every student tentatively held at $c$ retains her application under straightforward behavior. In each subsequent round, college $c$ selects its $q_c$ highest-priority students from a set containing all students held in the preceding round together with any newly arriving applicants. Hence, the minimum score among the students held at $c$, and therefore its cutoff, cannot decrease. 

\begin{remark} The benefit from skipping can be measured relative to DA in which a rejected student applies to the next untried college on her preference list, step by step. Suppose that, after a rejection, student $i$ has $m$ untried colleges ranked above her most-preferred currently attainable college. Without skipping, student $i$ would require $m+1$ rounds to reach the attainable college. Under TCDM, she applies to that college in the next round, thereby saving $m$ application rounds. By Lemma \ref{lemma:monotone}, none of the skipped colleges could have become attainable in  later rounds.
\end{remark}

We assume that students follow straightforward strategies under TCDM for two reasons. First, our aim is to analyze the effect of the deadline. Outcomes under TCDM depend both on the number of rounds and on how students play, thus fixing behavior allows us to attribute outcome differences to the deadline. Straightforward behavior is a natural benchmark that extends
truthful reporting under DA to the dynamic setting and is the strategy studied in the literature on sequential mechanisms. 
Second, this behavior assumption has empirical support. In experiments on sequential mechanisms, although without a time constraint,  a high proportion of subjects play this strategy \citep{BoHa2016,gong2016dynamic}. The survey we conducted following the 2022 college admissions process in Inner Mongolia also points to the use of straightforward strategies.\footnote{The survey was conducted on the day of the 2022 application and shortly afterward, at two high schools where students submitted applications from the schools' computer labs; COVID restrictions
limited the survey to these two schools. Of 850 students invited, 142 completed the survey, yielding 138 valid responses. More than 50
percent of respondents reported intending to use a straightforward
strategy, and 31 percent reported settling early on the best college they could reach and making no further revisions near the end of the process. Survey questions are provided in the Online Appendix.
} 

\subsection{When do straightforward strategies form an equilibrium?}\label{sec:eq}

For the equilibrium analysis in this subsection, we assume complete information: students' preferences and scores, colleges' capacities, and the number of rounds are common knowledge. The equilibrium properties of straightforward behavior depend  on whether the deadline leaves enough time for the application-and-rejection process to be completed. We first show that the final-round continuation game has a simple structure; we then analyze straightforward strategies separately for the cases $T\geq |I|$ and $T<|I|$.

For any point reached at the end of a round, we refer to the
\textit{continuation game} as the game induced by TCDM from that point onward, taking all preceding applications as fixed. At the end of Round $T-1$, the continuation game consists only of Round $T$: every student chooses a college or the outside option, and the final matching is determined by the resulting Round-$T$ application profile.

Consider first $T=1$, where the mechanism consists of a single
simultaneous application round. In this case, every Nash equilibrium outcome coincides with the unique stable matching. 
To see this, suppose students play some strategy profile where the outcome $\mu'$ differs from the stable matching. Since the stable matching is unique, $\mu'$ is unstable. Hence, either some student $i$ prefers the outside option to $\mu'(i)$, or there exists a student--college pair $(i,c)$ such that student $i$ prefers $c$ to $\mu'(i)$ and college $c$ either has a vacant seat or admits a lower-priority student. In the first case, student $i$ can profitably deviate by choosing the outside option; in the second case, she can profitably deviate by applying to $c$.

As a result, in every Nash equilibrium, each student assigned to a college under the stable matching applies to that college, while a student who is unassigned under the stable matching may choose the outside option or apply to any college at which she is rejected. Therefore, when $T=1$, a profile of straightforward strategies is an equilibrium if and only if  every student is either assigned to her top choice or not assigned in the stable matching.

When $T>1$, because the final matching depends only on the final-round application profile, the continuation game in the last round of the TCDM is, regardless of the preceding history, strategically identical to this one-round game. The remark below summarizes this observation. 

\begin{remark}
After any realized sequence of applications through the end of Round $T-1$, the continuation game is equivalent to the one-round application game, since every student can retain or revise her application and the final matching depends only on the Round-$T$ application profile. Every Nash equilibrium outcome of this continuation game coincides with the unique stable matching. Consequently, under complete information, every subgame-perfect equilibrium of TCDM induces the unique stable matching, regardless of the number of rounds.

\end{remark}

This observation is related to \citet{BoHa2016},  Proposition 3, who show that the Nash and subgame-perfect Nash equilibrium outcomes of the sequential mechanism inspired by the Brazilian college admissions coincide with the set of stable outcomes. Under our common-priority assumption, the stable matching is unique. Our formulation emphasizes that the same conclusion applies to the final-round continuation game after every preceding history.

\paragraph{The case of $T \geq \vert I\vert$.} When the maximum number of rounds is no less than the number of students,  \autoref{NTC} shows that straightforward play reaches the unique stable matching, while \autoref{NTC2} shows that the profile of straightforward strategies is a Nash equilibrium. Intuitively, the history defined by a profile of straightforward strategies closely resembles the steps of the DA algorithm. Students apply to their most preferred option in the first round and, after being rejected, apply to the most preferred option among those that are attainable. 

The bound of $\vert I\vert$ rounds required for straightforward play to reach the stable matching is tight. For example, suppose that all students have the same preferences, every college is preferred to the outside option by every student, all colleges have unit capacity, and $\vert C\vert\geq \vert I\vert$. In this case, exactly one additional student becomes assigned in each round, and the lowest-priority student is accepted in round $\vert I\vert$. Hence, the stable matching is reached after exactly $\vert I\vert$ rounds.
We formally establish these results below; the proofs are provided in \autoref{proofs_S3}.

\begin{proposition}\label{NTC}
Suppose that students follow straightforward strategies. If the process is
allowed to continue until no student revises her application, it reaches the
unique stable matching after at most $|I|$ rounds. Hence, if $T\geq |I|$,
the final outcome of TCDM coincides with the unique stable matching.
\end{proposition}

\begin{proposition}\label{NTC2}
Suppose $T \geq \vert I\vert$. Then a profile in which students follow straightforward strategies is a Nash equilibrium of the game induced by TCDM.
\end{proposition}

The convergence conclusion of \autoref{NTC} extends to settings with college-specific strict priority orders, such as multiple tie-breaking.\footnote{Here, straightforward behavior is defined using each college's own priority order: after a rejection, a student skips any college at which she would be rejected given the currently held applicants.} If tentatively accepted students retain their applications until rejected, straightforward play implements a compressed version of the student-proposing DA algorithm, in which students skip colleges that would reject them immediately; thus, when allowed to continue until no student changes her application, the process reaches the student-optimal stable matching. However, the bound $T\geq |I|$ relies on the existence of a common priority order and need not hold with college-specific priorities, since the application-and-rejection process may last for more than $|I|$ rounds. The equilibrium conclusion of \autoref{NTC2} also relies on the common priority order.

\paragraph{The case of $T < \vert I\vert$.}
For every $T < \vert I\vert$, there exists a problem for which the profile of straightforward strategies is not a Nash equilibrium. To see this, suppose that all students have the same preferences,
\[
c_1 \succ_i c_2 \succ_i \cdots \succ_i c_{|C|}
\succ_i \emptyset
\qquad\text{for every }i\in I,
\]
all colleges have unit capacity, and $\vert C \vert \geq T+1$. Order students according to the common priority ranking, with student $i_k$ having the $k$-th highest priority.

Under straightforward behavior, in Round 1 all students apply to $c_1$, and only student $i_1$ is tentatively assigned. In Round 2, all rejected students apply to $c_2$, and only student $i_2$ is tentatively assigned. Proceeding similarly, at the end of each Round $t\leq T$, student $i_t$ is tentatively assigned to $c_t$, while all students with lower priority are rejected. Hence, student $i_{T+1}$ is rejected in each of the first $T$ rounds and remains unassigned when the mechanism stops.

Student $i_{T+1}$ can profitably deviate by applying to $c_{T+1}$ in Round 1. Since no other student applies to that college in the first round, she is assigned to it and retains her application thereafter. Because $c_{T+1}\succ_{i_{T+1}}\emptyset$, this deviation is profitable. 
Therefore, the profile of straightforward strategies is not a Nash equilibrium. 

The preceding analysis clarifies the role of straightforward behavior in the remainder of the paper. When sufficiently many rounds are available, straightforward behavior reaches the stable matching and is itself consistent with equilibrium. When the deadline binds, however, straightforward behavior need not constitute an equilibrium. Our analysis below therefore does not interpret the outcomes generated by a binding deadline as equilibrium predictions under full strategic sophistication. Rather, we study how the deadline affects assignments when students follow straightforward strategies.


\section{Comparisons with DA}\label{with_const}

In this section, we compare the outcomes of TCDM with those of DA for the same preference profile. For each preference profile $\succ$, let $\mu^{TCDM}(\succ)$ denote the final matching generated by the $T$-round TCDM when all students follow the straightforward strategy. Let $\mu^{DA}(\succ)$ denote the matching generated by DA when students report the preference profile $\succ$.

\subsection{Ex-post comparisons}\label{expost}

At the individual level, we distinguish two main effects of the time constraint. The first is the \textit{direct time-constraint effect}. It arises when a student is rejected in the final round and has no opportunity to submit another application. She therefore remains unassigned under TCDM. If the process were allowed to
continue until reaching the DA outcome, she would either remain unassigned or obtain an acceptable college. The effect is therefore weakly negative relative to DA. We illustrate it in the example below.  
\medskip 

\begin{example}[\textbf{The direct time constraint effect}]\label{time1}
Consider four students and four colleges, each with one seat, and suppose that TCDM has $T=2$ rounds. Assume that every student prefers every college to the outside option. The priorities are such that $i_1$ has the highest priority, followed by $i_2$, $i_3$, and $i_4$. 
Students' preferences are as follows:

\[
\begin{matrix}
\succ_1: & c_1 &  c_2   &  c_3  & c_4  \\
\succ_2: & c_1 &  c_2   &  c_4  & c_3  \\
\succ_3: & c_2 &  c_3   &  c_1  & c_4  \\
\succ_4: & c_3 &  c_4   &  c_1  & c_2  \\
\end{matrix}
\]
\smallskip

We first look at the matching process under TCDM (see panel (a) of \autoref{tab:direct}).
In Round 1, students apply to their most-preferred colleges, and all students except $i_2$ are tentatively assigned to their first choices.
In Round 2, students observe the tentative matching. The tentatively assigned students retain their applications (which is indicated in the figure by letting them apply to the same choice this round). 
The only rejected student, $i_2$, observes that her second choice, $c_2$, is attainable. She therefore applies to $c_2$. Since this is the final round, $c_2$ tentatively holds $i_2$ and rejects $i_3$. $i_3$ remains unassigned because she has no further opportunity to revise her
application. Under DA (see panel (b) of \autoref{tab:direct}), the process continues to Step 3, when student $i_3$ applies to $c_3$ and displaces student $i_4$. Student $i_4$ then applies to her second choice, $c_4$, in Step 4. Thus, student $i_3$ is directly harmed by the time constraint: she is unassigned under TCDM, whereas the continuation of the process under DA assigns her to $c_3$. $\square$

\begin{figure}[h!]
\caption{The direct time constraint effect}\label{tab:direct}
\centering 
\begin{subtable}{0.4\linewidth}
\centering 
  $
  \begin{matrix}
   & \text{Round } 1   & \text{Round } 2 & \\
  i_1: & \dboxed{c_1} & \boxed{c_1} &   &   \\
  i_2: & {c_1} &  \boxed{c_2}  &    &   \\
  i_3: &  \dboxed{c_2}        &  c_2 &    &   \\
  i_4: & \dboxed{c_3} &     \boxed{c_3}         &    &   \\
  \end{matrix}
  $
  \medskip 
  
  \caption{TCDM with $T=2$}
\end{subtable}
\begin{subtable}{0.5\linewidth}
\centering 
  $
  \begin{matrix}
   & \text{Step } 1 & \text{Step } 2 & \text{Step } 3 & \text{Step } 4 \\
  i_1: & \dboxed{c_1} & \dboxed{c_1} & \dboxed{c_1} & \boxed{c_1} \\
  i_2: & c_1           & \dboxed{c_2} & \dboxed{c_2} & \boxed{c_2} \\
  i_3: & \dboxed{c_2} & c_2           & \dboxed{c_3} & \boxed{c_3} \\
  i_4: & \dboxed{c_3} & \dboxed{c_3} & c_3           & \boxed{c_4}
\end{matrix}
  $
  \medskip 
  \caption{DA}
\end{subtable}

  \smallskip 
  
  \begin{minipage}{0.9\textwidth}
  \footnotesize 
  \textit{Notes:} The dashed box indicates tentative assignment, whereas the solid box indicates final assignment. 
  \end{minipage}
\end{figure}
\end{example}

The second effect is the \textit{indirect time-constraint effect}. It arises when the deadline interrupts a rejection chain before it reaches a student who would otherwise be displaced. The deadline may therefore allow that student to retain a college that she prefers to her assignment under DA. From her perspective, the indirect effect is positive.

\begin{example}[\textbf{The indirect time constraint effect}]\label{time2}
Consider the same setting as in \autoref{time1}. Under TCDM, student~$i_4$ is assigned to her first choice, $c_3$. She benefits from the deadline because student $i_3$, who has higher priority, is rejected in the final round and has
no opportunity to continue the rejection chain. 
Under DA, the process continues. In Step 3, student $i_3$ applies to $c_3$ and displaces student $i_4$, who then applies to $c_4$ in Step 4. Hence, student $i_4$ receives $c_3$ under TCDM but $c_4$ under DA. $\square$
\end{example}
\medskip 

\begin{remark}
The preceding examples show that:
\begin{enumerate}
    \item TCDM need not be stable.
    \item TCDM may be Pareto inefficient, and DA may Pareto dominate it. To see that DA may Pareto dominate TCDM, modify \autoref{time1} by changing student $i_4$'s preferences to
\[
c_4 \succ_4 c_3 \succ_4 c_1 \succ_4 c_2 \succ_4 \emptyset.
\]
Then every student except $i_3$ receives the same assignment under both mechanisms, whereas $i_3$ is unassigned under TCDM and assigned to $c_3$ under DA. Hence, DA Pareto dominates TCDM.
    \item Moving from DA to TCDM may generate both winners and losers.
\end{enumerate}
\end{remark}
	

The next lemma sharpens the intuition behind the indirect time-constraint effect. It shows that a student can benefit from truncation only because some higher-priority student remains unassigned when the process stops.

\begin{lemma}\label{redis}
Suppose students follow straightforward strategies. If a student prefers her assignment under TCDM to her assignment under DA, then there exists a student with higher priority who is unassigned under TCDM.
\end{lemma}

Straightforward behavior is essential to the argument in \autoref{redis}. It ensures that whenever a student in the rejection chain is assigned under TCDM, she must prefer her TCDM assignment to her DA assignment. This allows the argument to continue to a student with still higher priority. Without straightforward behavior, the capacity argument still identifies a higher-priority student who is assigned to the relevant college under DA but not under TCDM, but it does not imply that the chain continues until reaching an unassigned student. The common-priority assumption is also essential for the conclusion of \autoref{redis}. With college-specific priorities, there is no single global ranking according to which each successive student in the rejection chain has higher priority than the preceding student. A rejection-chain interpretation may continue to apply, but the lemma's conclusion that an unassigned student has higher priority than the beneficiary does not extend directly.

\subsection{Ex-ante comparisons}
We now compare DA and TCDM from an ex-ante perspective. We fix an arbitrary probability distribution over preference profiles and compare the assignment lotteries induced by the two mechanisms. For each realized preference profile, students follow straightforward strategies under TCDM and report their
preferences truthfully under DA.

\begin{example}\label{exantecompare}
Consider four students and four colleges, each with one seat, and suppose that TCDM has $T=2$ rounds. Priorities are
\[
i_1\triangleright i_2\triangleright i_3\triangleright i_4.
\]
Every student prefers every college to the outside option. Students' preferences are drawn independently and uniformly from the set of $4!$ strict rankings. To compute the assignment probabilities for a given student, we fix an arbitrary preference for that student and enumerate the $(4!)^3$ possible
preference profiles of the other three students. For each profile, we compute the assignment under TCDM, assuming straightforward behavior, and under DA, assuming truthful reporting. The probability of each assignment rank is the fraction of profiles that generate that outcome. Table \ref{table1} presents the results. 

\begin{table}[htp]
\centering 
\begin{threeparttable}
  \caption{Probabilities of assignment under TCDM and DA}\label{table1}
  \small 
\begin{tabular}{lccccccccccc}
\toprule
  &   \multicolumn{5}{c}{TCDM (T=2)} & & \multicolumn{5}{c}{DA} \\
\cmidrule{2-6}
\cmidrule{8-12}
 Choice     & 1st & 2nd  &   3rd & 4th &  Unassigned &  & 1st & 2nd   & 3rd & 4th & Unassigned\\
\midrule
$i_1$ &  1    & 0    & 0 & 0 & 0               &  & 1    & 0    & 0 & 0 & 0 \\
$i_2$ &  0.750 & 0.250 & 0 & 0  & 0               &  & 0.750 & 0.250 & 0 & 0 & 0 \\
$i_3$ &  0.500   & 0.292 & 0.125	&0	&0.083    &  & 0.500 & 0.333 & 0.167  & 0  & 0 \\
$i_4$ &  0.271	&0.205 &	0.146	&0.094 & 0.285 &  & 0.250 &	0.250 & 0.250  & 0.250 & 0\\
\bottomrule
\end{tabular}
\begin{tablenotes}[flushleft]
\footnotesize
\item Notes: Entries are rounded to three decimal places.
\end{tablenotes}
\end{threeparttable}
\end{table}

First, the two-round deadline does not affect students $i_1$ and $i_2$. Because $i_1$ has the highest priority, she always receives her first choice under both mechanisms. Student $i_2$ has the second-highest priority, and her assignment is also identical under the two mechanisms for every preference profile. 

For student $i_3$, the probability of receiving her first choice is the same under both mechanisms. This is because the two students with higher priority receive the same assignments under TCDM and DA for every realized preference profile. However, student $i_3$ is less likely to receive any of her lower-ranked choices under TCDM. The reason is the direct negative effect: she may be rejected in the final round and remain unassigned, whereas under DA she would have an opportunity to continue applying. Consequently, the assignment distribution of student $i_3$ under DA first-order stochastically dominates her assignment distribution under TCDM.

Student $i_4$, by contrast, has a higher probability of receiving her first choice under TCDM. This reflects the indirect time-constraint effect: the deadline may interrupt a rejection chain before a higher-priority student displaces her. At the same time, student $i_4$ may experience the direct effect if she is rejected in the final round and has no opportunity to continue applying.$\square$

\end{example}
\smallskip

In the next proposition, we compare the distribution over ranked options of each student under TCDM and DA. For this, we need to define those students that are not constrained under TCDM.  

\begin{definition}\label{def:unconstrained}
For a given number of rounds $T$, student $i$ is \textbf{$T$-unconstrained} if
\[
\mu^{TCDM}(\succ)(i)=\mu^{DA}(\succ)(i)
\]
for every preference profile $\succ$. 
\end{definition}

Whenever we impose additional restrictions on preferences---for example, that every college is acceptable to every student---the quantification over preference profiles in \autoref{def:unconstrained} is understood relative to the corresponding restricted domain.

The preceding definition has a simple characterization under the common priority order. Relabel the colleges so that
\[
q_{c_1}\leq q_{c_2}\leq\cdots\leq q_{c_{|C|}},
\]
breaking ties arbitrarily. For each $k=1,\ldots,|C|$, define
\[
\Sigma(k):=\sum_{m=1}^k q_{c_m}.
\]
Thus, $\Sigma(k)$ is the combined capacity of the $k$ colleges with
the fewest seats.

\begin{lemma}\label{lem:unconstrained}
Suppose that students follow straightforward strategies. Moreover, assume that every college is acceptable to every student, that
\[
\sum_{c\in C}q_c\geq |I|,
\]
and $T\leq |C|$. Let $r_i$ denote student $i$'s position in the common priority order. Then student $i$ is $T$-unconstrained if and only if
\[
r_i\leq\Sigma(T).
\]
\end{lemma}

Before our next result, we need to introduce the following definition. Fix an arbitrary probability distribution $\mathbb{P}$ over preference profiles. For each preference profile $\succ$, let $c_i^l(\succ_i)$ denote student
$i$'s $l$-th-ranked college under $\succ_i$. Define
\[
p^{TCDM}_{i,l}
=
\mathbb{P}\!\left( 
\mu^{TCDM}(\succ)(i)=c_i^l(\succ_i)
\right)
\]
and
\[
p^{DA}_{i,l}
=
\mathbb{P}\!\left(
\mu^{DA}(\succ)(i)=c_i^l(\succ_i)
\right).
\]

For the outside option, define
\[
p^{TCDM}_{i,\emptyset}
=
\mathbb{P}\!\left(\mu^{TCDM}(\succ)(i)=\emptyset\right),
\qquad
p^{DA}_{i,\emptyset}
=
\mathbb{P}\!\left(\mu^{DA}(\succ)(i)=\emptyset\right).
\]

\begin{proposition}\label{prop: ex-ante prob}
Suppose that students follow straightforward strategies under TCDM and report their preferences truthfully under DA. Fix an arbitrary probability distribution over preference profiles. Then, for every student $i$:

\begin{enumerate}
    \item If student $i$ is $T$-unconstrained,
    \[
    p^{TCDM}_{i,l}=p^{DA}_{i,l}
    \qquad
    \text{for every }l\in\{1,\ldots,|C|\},
    \]
    and
    \[
    p^{TCDM}_{i,\emptyset}=p^{DA}_{i,\emptyset}.
    \]

    \item
    \[
    p^{TCDM}_{i,1}\geq p^{DA}_{i,1}.
    \]
\end{enumerate}

If, in addition, every student prefers every college to the outside option and
\[
\sum_{c\in C}q_c\geq |I|,
\]
then, for every student $i$:

\begin{enumerate}
    \setcounter{enumi}{2}

    \item
    \[
    p^{TCDM}_{i,\emptyset}\geq p^{DA}_{i,\emptyset}=0.
    \]

    \item
    \[
    \sum_{l=2}^{|C|}p^{TCDM}_{i,l}
    \leq
    \sum_{l=2}^{|C|}p^{DA}_{i,l}.
    \]
\end{enumerate}
\end{proposition}

The event inclusions and assignment equalities underlying \autoref{prop: ex-ante prob} hold profile by profile. Consequently, the probability comparisons are valid under any distribution over preference profiles, provided that both mechanisms are
evaluated under the same distribution. Also, the main comparisons in Proposition \ref{prop: ex-ante prob} extend to college-specific strict priorities. Under straightforward behavior, TCDM is a truncation of student-proposing DA, so $p_{i,1}^{TCDM}\geq p_{i,1}^{DA}$; moreover, with sufficient aggregate capacity and universal acceptability, $p_{i,\emptyset}^{TCDM}\geq p_{i,\emptyset}^{DA}=0$. Consequently, part 4 of the proposition also holds. What does not extend directly is the rank-based characterization in \autoref{lem:unconstrained}. With college-specific priorities, there is no single priority rank. The definition of a $T$-unconstrained student, based on equality between her TCDM and DA assignments for every preference profile, remains applicable.

The next corollary identifies a sufficient condition under which a student strictly prefers the assignment lottery generated by TCDM to that generated by DA. For the following utility comparison, fix student $i$'s preference ordering $\succ_i$ and interpret the assignment probabilities under a distribution over the other students' preferences, conditional on $\succ_i$.

\begin{corollary}\label{coro1}
Suppose that students follow straightforward strategies under TCDM and report their preferences truthfully under DA. Fix a student $i$ for whom every college is acceptable. If
\[
p^{TCDM}_{i,1}>p^{DA}_{i,1},
\]
then there exists a von Neumann--Morgenstern utility representation
consistent with her ordinal preferences for which she strictly prefers the outcome under TCDM to the outcome under DA.
\end{corollary}

The conclusion of \autoref{coro1} also extends to college-specific strict priorities. The conclusion only requires
\[
p^{TCDM}_{i,1}>p^{DA}_{i,1},
\]
since the utility of the first choice can be chosen sufficiently large to offset any losses in the probabilities of receiving lower-ranked colleges. 

The comparison in \autoref{coro1} should not be interpreted as a normative argument against stability. The corollary only shows that, when the first-choice advantage is strict, some cardinal utility representations make the student prefer the TCDM lottery. TCDM may nevertheless produce unstable outcomes and, under universal acceptability and sufficient aggregate capacity, gives the student a weakly higher probability of remaining unassigned.

In the final part of this subsection, we study how assignment probabilities
under TCDM change as the number of rounds increases.

\begin{proposition}\label{prop:rounds}
Suppose that students follow straightforward strategies. Fix an arbitrary probability distribution over preference profiles. For each student $i$, let $p_{i,l}^{TCDM(T)}$ denote the probability that she is assigned to her $l$-th choice under TCDM with $T$ rounds, and let
$p_{i,\emptyset}^{TCDM(T)}$ denote the probability that she remains unassigned. Then:

\begin{enumerate}
    \item For every student $i$,
    \[
    p_{i,1}^{TCDM(T)}\geq p_{i,1}^{TCDM(T+1)}.
    \]

    \item The expected number of students assigned to a non-first-choice college weakly increases from $T$ to $T+1$:
    \[
    \sum_{i\in I}\sum_{l=2}^{|C|}p_{i,l}^{TCDM(T)}
    \leq
    \sum_{i\in I}\sum_{l=2}^{|C|}p_{i,l}^{TCDM(T+1)}.
    \]
\end{enumerate}
\end{proposition}

Both conclusions of \autoref{prop:rounds} continue to hold under arbitrary strict college-specific priorities. Under straightforward behavior, a student who receives her first choice after $T+1$ rounds must also have been tentatively assigned to it after Round $T$, since rejection from her first choice is final. Hence,
\[
p_{i,1}^{TCDM(T)}\geq p_{i,1}^{TCDM(T+1)}
\]
for every student.

Likewise, an additional round weakly increases the number of occupied seats and weakly decreases the number of students assigned to their first choices. Therefore,
\[
\sum_{i\in I}\sum_{l=2}^{|C|}p_{i,l}^{TCDM(T)}
\leq
\sum_{i\in I}\sum_{l=2}^{|C|}p_{i,l}^{TCDM(T+1)}.
\]

\subsection{College-specific eligibility}\label{subsec:eligibility}

The baseline model assumes that every student is eligible for admission to every college. We now consider an extension in which colleges may impose different eligibility criteria. Let $\triangleright_0$ denote the common priority order. Each college $c\in C$ has a set of eligible students
\[
A_c\subseteq I.
\]
Priorities among students who are eligible for the same college are aligned with the common priority order: for every $i,j\in A_c$,
\[
i\triangleright_c j
\quad\Longleftrightarrow\quad
i\triangleright_0 j.
\]
Thus, colleges may differ in their eligibility requirements, but any two students who are eligible for the same college are ranked according to $\triangleright_0$. A student outside $A_c$ is rejected by college $c$ regardless of its remaining capacity or current cutoff. We assume that the eligibility sets are common knowledge.

Under college-specific eligibility, a student--college pair $(i,c)$ can block a matching only if $i\in A_c$. Thus, an ineligible student cannot form a blocking pair with college $c$.

Under this extension, straightforward behavior is modified as follows. In Round~1, student $i$ applies to her most-preferred option in
\[
\{c\in C:i\in A_c\}\cup\{\emptyset\}.
\]
In every subsequent round, a tentatively assigned student retains her current application. A student who is unassigned after Round $t-1$ applies to her most-preferred option in
\[
\{c\in C:i\in A_c\text{ and }s_i\geq f_c^{t-1}\}
\cup\{\emptyset\},
\]
where $f_c^{t-1}$ denotes the cutoff of college $c$ after Round $t-1$.

The main structural conclusions of \autoref{sec:eq} and the rejection-chain result in \autoref{expost} continue to hold. The stable matching remains unique and coincides with the outcome of serial dictatorship following $\triangleright_0$, subject to college-specific eligibility. 

Under the modified notion of straightforward behavior,
\autoref{NTC} continues to hold. Ineligibility merely removes some colleges from a student's set of feasible applications. A student with the $k$-th position in the common priority order can still be displaced only by one of the $k-1$ students with a higher priority than her. Hence, by the end of Round $k$, she has reached her serial-dictatorship assignment, and the process reaches the unique stable matching after at most $|I|$ rounds.

The equilibrium conclusion in \autoref{NTC2} also continues to hold when $T\geq |I|$. A unilateral deviation cannot enable a student to obtain a college for which she is ineligible. Moreover, every eligible college that she prefers to her stable assignment is filled, under serial dictatorship, by students with higher priority in the common order. Since her deviation cannot alter the assignments of these higher-priority students, she cannot obtain a strictly preferred assignment.
The rejection-chain conclusion in \autoref{redis} extends for the same
reason. 

College-specific eligibility affects some of the ex-ante comparisons. The first-choice comparison
\[
p_{i,1}^{TCDM}\geq p_{i,1}^{DA}
\]
continues to hold (when ``first choice'' is interpreted as the student's most-preferred eligible college), as do both conclusions of \autoref{prop:rounds}, with choice ranks defined over the student's eligible colleges. The definition of a $T$-unconstrained student also remains applicable. However, the rank-based characterization in
\autoref{lem:unconstrained} need not hold, because the colleges available to a student now depend on her eligibility.

Moreover, sufficient aggregate capacity no longer guarantees that DA assigns every student. Eligibility restrictions may leave some students unassigned
even when some colleges have vacant seats. Consequently, the comparison in Parts~3 and~4 of \autoref{prop: ex-ante prob} need not extend
without additional assumptions. 

\section{Conclusions}\label{sec: conclusion}
Advances in information technology have led school-choice and college-admission systems to adopt dynamic procedures in which students can revise their applications in response to feedback. These procedures differ from traditional one-shot application systems and may allow students to react more effectively to their
evolving admission prospects. Our analysis shows, however, that an exogenous deadline can materially alter their outcomes.

We study dynamic college-admission procedures with tentative acceptances and a fixed number of rounds, and compare their outcomes with those of student-proposing DA. Under a common priority order, when sufficiently many rounds are available, straightforward behavior reaches the unique stable matching and, under complete information, constitutes a Nash equilibrium. When the deadline binds, by contrast, it may interrupt rejection chains before they are completed. This can directly harm students who are rejected in the final round and remain unassigned, while indirectly benefiting students who retain colleges from which they would eventually be displaced under DA.

The ex-ante comparisons reveal a related trade-off. TCDM weakly increases each student's probability of receiving her first choice relative to DA. Under universal acceptability and sufficient aggregate capacity, however, it also weakly increases the probability of remaining unassigned and weakly reduces each student's probability of receiving a non-first-choice college. As the number of rounds increases, each student's probability of receiving her first choice weakly decreases, while the expected number of students assigned to
non-first-choice colleges weakly increases. Thus, allowing additional rounds need not improve every student's assignment.

One challenge in utilizing field data generated by the dynamic mechanism is that we do not observe the students' actual preferences. Controlled laboratory experiments, in which preferences can be induced or directly observed, could help quantify more precisely the benefits and costs of binding deadlines. Information provision is another promising design
margin. In the Inner Mongolia procedure, all students receive the same type
of information, although the value of that information may differ across students. More targeted feedback could accelerate the revision process and mitigate the consequences of a binding deadline.

\bibliographystyle{econometrica}
\bibliography{Mongolia}

\appendix 

\newpage 
\section*{Appendix}

\section{Data}\label{data}
The Inner Mongolia procedure provides two types of information. The first type is updated in real time and is privately available to each student. It reports the cutoff score at the university to which the student
has applied, her rank among applicants to that university, and her rank among applicants to her first-choice program. The second, updated hourly, shows the cutoff scores of all universities, the applicants to each university and the tentative assignments. The hourly updated information is made public and we use it to construct a dataset to study students’ application behavior and tentative assignment outcomes over time. Since the hourly data are provided at the university level and lack unique student identifiers, we link observations across hours using the procedure below.

First, we leverage the staggered closing rule. In the last hour, only students from the final batch can revise their choices. In the previous hour, these students, along with those from an earlier batch, can revise their applications, while the rest remain inactive. This allows us to precisely link the inactive students across hours. We refer to a linked pair as a \textit{link}. 

Second, for students who are active in both the last and previous hours, common characteristics may not uniquely identify them. To resolve this, we apply two rules below.

\textbf{Rule 1: Program-level matching.} At a given university, if there is a unique observation with the same characteristics and program choices as another observation in a previous hour, we treat them as the same student, assuming that the student has not changed either their program or university choice.

\textbf{Rule 2: University-level matching.}
 At a given university, if there is a unique observation with the same characteristics as another student in a previous hour, we treat them as the same student, assuming no change in university choice.

The first rule essentially treats two students observed at different hours as the same individual if they share the same characteristics, university, and program choices. After initially linking students using the program-level matching, we apply the second rule to link those who may have changed their program choices but not their university choice. If no match is found, we then search for another student with the same characteristics at a different university in an earlier hour.

Applying this Java-based procedure to the 2018 university-level hourly data, we linked all students in the final-hour data---21,107 from the STEM track and 6,089 students from the humanities track---to records in previous hours.

\section{The extensive-form game induced by TCDM}\label{app: extensive}

We describe the game induced by TCDM in this section. The definitions largely follow \citet{klijn2019static}. Throughout this appendix, we fix an admissions problem and assume complete information: students' preferences and scores, colleges' capacities and
priorities, and the number of rounds are common knowledge. The game has the following elements.

\begin{itemize}
    \item The set of players is
\[
I=\{1,\ldots,n\}.
\]

\item A common action set
\[
A\equiv C\cup\{\emptyset\}.
\]
At each decision node, a student chooses a college to which to apply or the outside option $\emptyset$. We adopt the convention that, if a student does not revise her application in a given round, she chooses the same action as in the preceding round.
    
\item A set of histories defined as

$$H:=\bigcup_{m=0}^{Tn}A^m,
\qquad
A^0:=\{h_0\},
\qquad
E:=A^{Tn}.$$

A history records the sequence of applications submitted up to a given
decision node. Each block of $n$ actions constitutes the application profile
in one round. The first $n$ actions constitute the application profile in Round~1, the next $n$ actions constitute the application profile in Round~2, and so on. Thus, the action of student $i$ in Round $t$ appears in position
\[
(t-1)n+i.
\]      
The set $E$ is the set of terminal histories.

\item A function $\iota : H \setminus E \rightarrow I$ that indicates whose turn it is to move at each decision node in $H \setminus E$.

We sequentialize students' actions within each round only to represent their simultaneous choices in extensive form. Students do not observe the actions taken by other students earlier in the same round. Student $1$ moves first, $2$ moves second, and so on. Thus, the player function is
$$\iota(h):=(|h|\bmod n)+1
\qquad
\text{for every }h\in H\setminus E,$$

where $|h|$ is the length of the sequence $h$.

\item For each student $i$, an information partition $\mathcal I_i$ of the set
\[
\{h\in H\setminus E:\iota(h)=i\}.
\]
Consider two histories $h$ and $h'$ at which student $i$ moves in the same round. Then $h$ and $h'$ belong to the same information set if and only if their first
\[
|h|-(i-1)
\]
components coincide. Thus, when student $i$ applies in a given round, she observes all applications submitted in previous rounds, but not the applications submitted by other students in the current round.

\item For any history $h$ reached at the end of Round $k$, so that
\[
|h|=kn
\qquad
\text{for some }k\in\{1,\ldots,T\},
\]
let $\mu^h$ denote the tentative matching induced by the applications
submitted in Round $k$. More precisely, if
\[
h=(a_1,\ldots,a_{kn}),
\]
then student $i$'s application in round $k$ is
\[
a_{(k-1)n+i},
\qquad i\in I,
\]
and $\mu^h$ is the matching obtained from the application profile
\[
\left(a_{(k-1)n+1},\ldots,a_{kn}\right)
\]
according to colleges' capacities and priorities. The outcome function
\[
g:E\rightarrow\mathcal M
\]
is defined by
\[
g(h):=\mu^h
\qquad
\text{for every }h\in E,
\]
where $\mathcal M$ denotes the set of feasible matchings.
\end{itemize}

For any $h,h'\in E$,
\[
h\succeq_i^E h'
\quad\Longleftrightarrow\quad
g(h)(i)\succeq_i g(h')(i).
\]

\begin{definition}
A pure strategy for student $i$ is a function
\[
\sigma_i:\mathcal I_i\rightarrow A
\]
that assigns an action to every information set at which student $i$ moves.
\end{definition}

Let $H_i:=\{h\in H\setminus E:\iota(h)=i\}$. For every $h\in H_i$, let $I_i(h)\in\mathcal I_i$ denote the information set containing $h$.

Consider a history $h\in H_i$ at which student $i$ moves in Round $t>1$.
Then
\[
|h|=(t-1)n+i-1.
\]
If
\[
h=(a_1,\ldots,a_{(t-1)n+i-1}),
\]
let
\[
h^-:=(a_1,\ldots,a_{(t-1)n})
\]
denote the history at the end of Round $t-1$, and define
\[
\bar\mu(h):=\mu^{h^-}.
\]
Thus, $\bar\mu(h)$ is the tentative matching at the end of the round
preceding the one in which history $h$ occurs.

Because histories in the same information set have identical completed-round application profiles, $\bar\mu(h)$ is constant on each information set. Hence, the strategy defined below is well defined as a function of information
sets.

\begin{definition}
We say that student $i$ follows a straightforward strategy if, for every $h\in H_i$,
\[
\sigma_i\bigl(I_i(h)\bigr)
=
\begin{cases}
\displaystyle
\max_{\succ_i}\bigl(C\cup\{\emptyset\}\bigr),
& \text{if $h$ occurs in Round~1,}\\[1.2ex]
\bar\mu(h)(i),
& \text{if $h$ occurs in a Round $t>1$ and }
  \bar\mu(h)(i)\neq\emptyset,\\[1.2ex]
\displaystyle
\max_{\succ_i}
\left(
\{c\in C:s_i\geq f_c({\bar\mu(h)})\}
\cup\{\emptyset\}
\right),
& \text{if $h$ occurs in a Round $t>1$ and }
  \bar\mu(h)(i)=\emptyset.
\end{cases}
\]
\end{definition}

Thus, in Round~1, a straightforward student applies to her most-preferred option. In every subsequent round, a tentatively assigned student retains her current application. An unassigned student applies to her most-preferred option among the outside option and the colleges whose observed cutoffs do not exceed her score. 

\setcounter{figure}{0} \renewcommand{\thefigure}{C.\arabic{figure}}

\section{Omitted proofs} \label{proofs_S3}

\paragraph{Proof of \autoref{NTC}.}
\begin{proof}
First, suppose that the process reaches a round after which no student changes her application. We show that the resulting tentative matching $\mu$ is stable.

Individual blocking cannot occur. Under straightforward behavior, the outside option is always attainable, so no student applies to a college that she ranks below $\emptyset$.

Suppose, toward a contradiction, that a student--college pair $(i,c)$ blocks $\mu$. Then
\[
c\succ_i\mu(i)
\]
and either college $c$ has a vacant seat or student $i$ has higher priority than some student assigned to $c$. This implies
\[
s_i>f_c(\mu).
\]

Consider first the case in which student $i$ is unassigned. Since $c$ is acceptable and currently attainable to her, straightforward behavior requires her to apply in the following round to an option that she weakly prefers to $c$, contradicting the assumption that no student changes her application.

Now suppose that student $i$ is tentatively assigned. Because
$c\succ_i\mu(i)$, before applying to her current college she must either have been rejected from $c$ or have skipped $c$ because its cutoff exceeded her score. Cutoffs weakly increase under straightforward behavior. Hence, the current cutoff of $c$ must be at least as high as it was when student $i$ was rejected from or skipped $c$. Therefore,
\[
s_i\leq f_c(\mu),
\]
contradicting the fact that $(i,c)$ blocks $\mu$. Thus, $\mu$ is stable and, by uniqueness, equals the unique stable matching.


It remains to establish the bound. Order students according to the common priority order as
\[
i_1\triangleright i_2\triangleright\cdots\triangleright i_{|I|}.
\]
Let $\mu^S$ denote the unique stable matching, equivalently the outcome of serial dictatorship following this order. We prove by induction that, by the end of Round $k$, every student $i_1,\ldots,i_k$ has reached her assignment under $\mu^S$ and will never move again.

For $k=1$, student $i_1$ applies to her most-preferred option in Round 1 and cannot be displaced, so the claim holds.

Suppose that it holds for the first $k-1$ students. By the beginning of Round $k$, these students permanently occupy their serial-dictatorship assignments. Every college that student $i_k$ prefers to $\mu^S(i_k)$ is filled, under serial dictatorship, by students among $i_1,\ldots,i_{k-1}$. Hence, by the induction hypothesis, each such college is unattainable to $i_k$ by Round $k$. 

We now distinguish two cases. First, suppose that
\[
\mu^S(i_k)=\emptyset.
\]
Then every college that student $i_k$ prefers to the outside option is filled under serial dictatorship by students among $i_1,\ldots,i_{k-1}$. By the induction hypothesis, these colleges are already permanently occupied by the beginning of Round $k$. By the beginning of Round $k$, student $i_k$ has been displaced from any acceptable college to which she may previously have applied. She is therefore unassigned and, in Round $k$, chooses the outside option, thereby reaching her
serial-dictatorship assignment.

Second, suppose that
\[
\mu^S(i_k)\neq\emptyset.
\]
The college $\mu^S(i_k)$ has capacity remaining after the assignments of $i_1,\ldots,i_{k-1}$. Therefore, if student $i_k$ is unassigned at the beginning of Round $k$, straightforward behavior makes her apply to $\mu^S(i_k)$, and no lower-priority student can prevent her admission.

If student $i_k$ is already tentatively assigned, her assignment must be $\mu^S(i_k)$. She cannot be assigned to a more-preferred college because those colleges are occupied by higher-priority students. She also cannot be
assigned to a less-preferred college. Indeed, since cutoffs weakly increase, if $\mu^S(i_k)$ is attainable after the higher-priority students have settled, it must also have been attainable whenever $i_k$ was previously unassigned. Straightforward behavior would therefore not have led her to apply to a less-preferred college. Thus, in either case, student $i_k$ reaches $\mu^S(i_k)$ by the end of
Round $k$ and never subsequently moves. This completes the induction step.

The induction implies that all students have reached their stable assignments by the end of Round $|I|$. Hence, the unique stable matching is reached after at most $|I|$ rounds.
\end{proof}

\paragraph{Proof of \autoref{NTC2}.}
\begin{proof}
Fix a student $i$, and suppose that every student other than $i$ follows a
straightforward strategy. We show that student $i$ has no profitable
deviation.

A deviation by student $i$ cannot affect the application paths or assignments of students with higher priority than $i$. Indeed, student $i$ cannot displace a higher-priority student. Moreover, although her application may change a college's cutoff, it cannot raise that cutoff above the score of any higher-priority student. Consequently, it does not change which colleges are attainable to those students under straightforward behavior. 

By \autoref{NTC}, when $T\geq |I|$, those higher-priority students reach the assignments they receive in the unique stable matching, equivalently in serial dictatorship following the common priority order. Under serial dictatorship, every college that student $i$ prefers to her stable assignment is filled to capacity by higher-priority students. Since these higher-priority students retain the same assignments when only student $i$ deviates, student $i$ cannot obtain any college that she prefers to her straightforward-strategy assignment.

Therefore, student $i$ has no profitable deviation. Since the argument applies to every student, the profile of straightforward strategies is a Nash equilibrium.
\end{proof}

\paragraph{Proof of \autoref{redis}.}

\begin{proof}
Fix a preference profile $\succ$ and suppress it from the notation. Suppose that student $i_0$ prefers her TCDM assignment to her DA assignment, and let
\[
c_0:=\mu^{TCDM}(i_0).
\]
Because
\[
c_0\succ_{i_0}\mu^{DA}(i_0),
\]
student $i_0$ applies to $c_0$ under DA and is rejected. Hence, $c_0$ is
filled under DA by students with higher priority than $i_0$.

Since student $i_0$ is assigned to $c_0$ under TCDM but not under DA, there
exists a student
\[
i_1\in
\bigl(\mu^{DA}\bigr)^{-1}(c_0)
\setminus
\bigl(\mu^{TCDM}\bigr)^{-1}(c_0).
\]
Moreover,
\[
i_1\triangleright i_0.
\]
If $i_1$ is unassigned under TCDM, the proof is complete.

Suppose instead that $i_1$ is assigned under TCDM, and let
\[
c_1:=\mu^{TCDM}(i_1).
\]
We claim that
\[
c_1\succ_{i_1}c_0.
\]
Indeed, if student $i_1$ preferred $c_0$ to $c_1$, then under straightforward behavior she would have applied to $c_0$ before applying to $c_1$, unless the cutoff of $c_0$ already exceeded her score. In either case, because cutoffs weakly increase under straightforward behavior (\autoref{lemma:monotone}), the lower-priority student $i_0$ could not subsequently be assigned to $c_0$. This contradicts $\mu^{TCDM}(i_0)=c_0$. Therefore,
\[
c_1\succ_{i_1}c_0=\mu^{DA}(i_1).
\]

Student $i_1$ consequently applies to $c_1$ under DA and is rejected. By the
same capacity argument, there exists
\[
i_2\in
\bigl(\mu^{DA}\bigr)^{-1}(c_1)
\setminus
\bigl(\mu^{TCDM}\bigr)^{-1}(c_1)
\]
such that
\[
i_2\triangleright i_1.
\]
If $i_2$ is unassigned under TCDM, the proof is complete. Otherwise, the same
argument can be repeated.

This procedure generates a sequence
\[
i_{m+1}\triangleright i_m
\qquad\text{at every step.}
\]
At every step, if the current student is assigned under TCDM, the preceding argument constructs another student with strictly higher priority. Because the common priority order is finite, this construction cannot continue indefinitely. It must therefore terminate at a student who is unassigned under
TCDM. Since priority strictly increases at every step, this student has higher priority than $i_0$, completing the proof.\end{proof}

\paragraph{Proof of \autoref{lem:unconstrained}.}
\begin{proof}
Suppose first that $r_i\leq\Sigma(T)$. Fix an arbitrary preference profile.
We begin with the following claim.

\medskip

\noindent\textit{Claim.}
If a student $j$ is unassigned after Round $t$, then there exist $t$
distinct colleges such that, after Round $t$, each of them is filled to
capacity with students who have higher priority than $j$. Consequently,
\[
r_j>\Sigma(t).
\]

\medskip
First observe that an unassigned student always has an attainable college. Otherwise, every college would be filled to capacity by students with higher priority than her, which would require at least
\[
\sum_{c\in C} q_c \geq |I|
\]
students with higher priority, an impossibility. Since every college is acceptable, an unassigned student therefore applies to a college rather than choosing the outside option.

We prove the claim by induction on $t$. Suppose first that $t=1$, and let $c$ be the college to which student $j$ applies in Round~1. If student $j$ is unassigned after Round~1, college $c$ must be filled to capacity with students who have higher priority than $j$. Since every college has capacity at least $\Sigma(1)$,
\[
r_j-1\geq q_c\geq\Sigma(1),
\]
and hence $r_j>\Sigma(1)$.

Now suppose that the claim holds for Round $t-1$, and let student $j$ be unassigned after Round $t$. Let $c$ be the college from which she is rejected in Round $t$. We distinguish two cases.

First, suppose that student $j$ was already unassigned after Round $t-1$. By the induction hypothesis, there exist $t-1$ distinct colleges\[c_1,\ldots,c_{t-1}\]that are filled after Round $t-1$ with students who have higher priority than $j$. Because cutoffs weakly increase, these colleges remain filled after Round $t$ with students who have higher priority than $j$. Moreover, none of these colleges is attainable to $j$ in Round $t$. Since $j$ applies to $c$,
\[c\notin\{c_1,\ldots,c_{t-1}\}.\]
Because $j$ is rejected from $c$, college $c$ is also filled after Round $t$ with students who have higher priority than $j$. 

Second, suppose that student $j$ was tentatively assigned after Round $t-1$. She then retains her application to $c$ in Round $t$. Her rejection must therefore be caused by a newly arriving student $j'$ with higher priority than $j$. Since $j'$ is newly arriving, she was unassigned after Round $t-1$. By the induction hypothesis, there exist $t-1$ distinct colleges\[c_1,\ldots,c_{t-1}\]that are filled after Round $t-1$ with students who have higher priority than $j'$, and therefore also higher priority than $j$. These colleges remain so after Round $t$. Furthermore, none of them is attainable to $j'$ in Round $t$. Since $j'$ applies to $c$,\[c\notin\{c_1,\ldots,c_{t-1}\}.\]

Finally, because $j$ is rejected from $c$, college $c$ is filled after Round $t$ with students who have higher priority than $j$. In either case, we have identified $t$ distinct colleges that are filled to capacity after Round $t$ with students who have higher priority than $j$. Their combined capacity is at least $\Sigma(t)$. Hence,\[r_j-1\geq \Sigma(t),\]and therefore\[r_j>\Sigma(t).\]

This proves the claim.

We now return to student $i$. Since
\[
r_i\leq\Sigma(T),
\]
the claim implies that student $i$ cannot be unassigned after Round $T$.

Let
\[
c^*:=\mu^{DA}(\succ)(i)
\]
be her DA assignment. Whenever student $i$ is unassigned, $c^*$ is attainable to her. Otherwise
its cutoff would be higher than $s_i$ at that history and, by cutoff
monotonicity, would remain so thereafter, contradicting \autoref{NTC},
according to which the continued process eventually assigns $i$ to $c^*$. Therefore, whenever student $i$ is unassigned, straightforward behavior never makes her apply to
a college that she ranks below $c^*$. It follows that
\[
\mu^{TCDM}(\succ)(i)\succeq_i\mu^{DA}(\succ)(i).
\]

Suppose, toward a contradiction, that the two assignments differ. Because
preferences are strict,
\[
\mu^{TCDM}(\succ)(i)\succ_i\mu^{DA}(\succ)(i).
\]
By \autoref{redis}, there then exists a student $j$ with higher priority than
$i$ who is unassigned under TCDM. Hence,
\[
r_j<r_i\leq\Sigma(T).
\]
This contradicts the claim, which requires every student who is unassigned
after Round $T$ to satisfy
\[
r_j>\Sigma(T).
\]
Therefore,
\[
\mu^{TCDM}(\succ)(i)=\mu^{DA}(\succ)(i).
\]

Since the preference profile was arbitrary, student $i$ is
$T$-unconstrained.

Conversely, suppose that $i$ is $T$-unconstrained, and by contradiction, that $r_i>\Sigma(T)$. Consider a preference profile in which all students rank the colleges as
\[
c_1\succ c_2\succ\cdots\succ c_{|C|}\succ\emptyset.
\]
After Round $T$, the first $T$ colleges jointly hold the top $\sum(T)$ students in the common priority order. Hence, student $i$ is unassigned under the $T$-round TCDM. Since every college is acceptable and aggregate capacity is sufficient, DA eventually assigns student $i$ to a college. Therefore, her assignments under the two mechanisms differ for this preference profile, so she is not $T$-unconstrained.
\end{proof}

\paragraph{Proof of \autoref{prop: ex-ante prob}.}

\begin{proof}
To simplify notation, fix a student $i$ and omit the subscript $i$.

Part~1 follows directly from \autoref{def:unconstrained}. 
For Part~2, fix a preference profile. Suppose $i$ receives her first choice under DA but is rejected from it under the $T$-round TCDM. Continue TCDM beyond Round $T$ under straightforward behavior. Since rejection from a college is final, $i$ can never return to
her first choice. \autoref{NTC}, however, implies that the continued process converges to the DA matching, where $i$ receives that college -- a
contradiction. Hence, student $i$ also remains assigned to her first choice under the $T$-round mechanism. Therefore,
\[
\left\{
\succ:\mu^{DA}(\succ)(i)
\text{ is $i$'s first choice}
\right\}
\subseteq
\left\{
\succ:\mu^{TCDM}(\succ)(i)
\text{ is $i$'s first choice}
\right\}.
\]
Taking probabilities yields
\[
p^{TCDM}_1\geq p^{DA}_1.
\]

For Part~3, because every college is acceptable to every student and
\[
\sum_{c\in C}q_c\geq |I|,
\]
DA assigns every student.
Thus,
\[
p^{DA}_{\emptyset}=0,
\]
and consequently
\[
p^{TCDM}_{\emptyset}\geq p^{DA}_{\emptyset}=0.
\]

Finally, under either mechanism, the student receives her first choice, a non-first-choice college, or the outside option. Hence,
\[
p^{TCDM}_1+\sum_{l=2}^{|C|}p^{TCDM}_l+p^{TCDM}_{\emptyset}=1,
\]
whereas
\[
p^{DA}_1+\sum_{l=2}^{|C|}p^{DA}_l=1.
\]
Using Parts~2 and~3,
\begin{align*}
\sum_{l=2}^{|C|}p^{TCDM}_l =1-p^{TCDM}_1-p^{TCDM}_{\emptyset} \leq 1-p^{DA}_1 =\sum_{l=2}^{|C|}p^{DA}_l.
\end{align*}
This proves Part~4.

\end{proof}

\paragraph{Proof of \autoref{coro1}.}

\begin{proof}
Consider a student such that
\[
p_{i,1}^{TCDM}>p_{i,1}^{DA}.
\]
Let $u_j$ denote the utility of being assigned to her $j$-th choice, and normalize the utility of being unassigned to zero. Fix any utilities satisfying
\[
u_2>u_3>\cdots>u_{|C|}>0.
\]
The difference in expected utility between TCDM and DA is
\[
EU^{TCDM}-EU^{DA}
=
\left(p_{i,1}^{TCDM}-p_{i,1}^{DA}\right)u_1
-
\sum_{j=2}^{|C|}
\left(p_{i,j}^{DA}-p_{i,j}^{TCDM}\right)u_j.
\]
Since $p_{i,1}^{TCDM}-p_{i,1}^{DA}>0$, choose $u_1$ such that
\[
u_1>
\max\left\{
u_2,\,
\frac{
\sum_{j=2}^{|C|}
\left(p_{i,j}^{DA}-p_{i,j}^{TCDM}\right)u_j
}{
p_{i,1}^{TCDM}-p_{i,1}^{DA}
}
\right\}.
\]
Then $u_1>u_2>\cdots>u_{|C|}>0$, so the utility vector represents the
student's ordinal preferences, and
\[
EU^{TCDM}>EU^{DA}.
\]
\end{proof}

\paragraph{Proof of \autoref{prop:rounds}.}

\begin{proof}
For Part 1, fix a preference profile. If student $i$ receives her first
choice after $T+1$ rounds, then she must also be tentatively assigned to it after Round $T$. Indeed, she applies to her first choice in Round 1, and, once rejected from it, she never applies to it again under straightforward behavior. Therefore,
\[
\left\{
\succ:\mu^{TCDM(T+1)}(\succ)(i)
\text{ is $i$'s first choice}
\right\}
\subseteq
\left\{
\succ:\mu^{TCDM(T)}(\succ)(i)
\text{ is $i$'s first choice}
\right\}.
\]
Taking probabilities yields
\[
p_{i,1}^{TCDM(T)}\geq p_{i,1}^{TCDM(T+1)}.
\]

For Part~2, fix a preference profile $\succ$, and let
\[
N^T(\succ)
:=
\left|
\left\{
i\in I:
\mu^{TCDM(T)}(\succ)(i)
\text{ is a non-first-choice college}
\right\}
\right|.
\]
We show that
\[
N^{T+1}(\succ)\geq N^T(\succ).
\]

At each college, every student assigned there after Round~$T$ retains her application in Round~$T+1$. If some of these students are displaced, at least as many newly arriving students are accepted in their place. Every newly arriving student in Round $T+1$ is applying to a non-first-choice college. Indeed, any student who ever applies to a college after Round~1 must initially have applied to her first-choice college and
subsequently been rejected from it; once rejected, she never applies to it again. Therefore, the number of students assigned to non-first-choice colleges cannot decrease. Hence,
\[
N^{T+1}(\succ)\geq N^T(\succ)
\]
for every preference profile $\succ$.

Taking expectations over preference profiles gives
\[
\mathbb{E}\!\left[N^{T+1}(\succ)\right]
\geq
\mathbb{E}\!\left[N^T(\succ)\right].
\]

Since
\[
\mathbb{E}\!\left[N^T(\succ)\right]
=
\sum_{i\in I}\sum_{l=2}^{|C|}
p_{i,l}^{TCDM(T)},
\]
and similarly
\[
\mathbb{E}\!\left[N^{T+1}(\succ)\right]
=
\sum_{i\in I}\sum_{l=2}^{|C|}
p_{i,l}^{TCDM(T+1)},
\]
we obtain
\[
\sum_{i\in I}\sum_{l=2}^{|C|}
p_{i,l}^{TCDM(T)}
\leq
\sum_{i\in I}\sum_{l=2}^{|C|}
p_{i,l}^{TCDM(T+1)}.
\]
\end{proof}

\newpage

\begin{center}
\textbf{Online Appendix}
\end{center}

\noindent \textit{Note: The original Chinese version was administered to respondents, and the English translation is provided for reference. 
}

\vspace{0.5cm}
\noindent Dear participant,

We are a research team from Beijing Normal University conducting a study on the college admission system. To better understand the application process in Inner Mongolia, we kindly ask you to spend about 10 minutes completing this questionnaire. Thank you for your participation!

All data collected will be used solely for research purposes. We strictly adhere to data regulations and ensure complete anonymity; your personal information will not be disclosed. Your responses will not affect your admission results in any way. If you have any questions regarding this survey or the research, please contact us at \href{mailto:zhu@bnu.edu.cn}{zhu@bnu.edu.cn}.

\newcounter{question}
\newcommand{\questionitem}{\stepcounter{question}\item[Q\thequestion.]}

\section*{Part 1: Basic Information}

\begin{enumerate}[leftmargin=*]
    \questionitem Your total Gaokao score: \underline{\hspace{4cm}} points.
    \questionitem Which study track do you belong to:
    \begin{enumerate}[label=(\alph*)]
        \item Humanities/Social Science 
        \item Science
        \item Other track
    \end{enumerate}
    \questionitem Your gender:
    \begin{enumerate}[label=(\alph*)]
        \item Female
        \item Male
    \end{enumerate}
    \questionitem Are you from an ethnic minority?
    \begin{enumerate}[label=(\alph*)]
        \item Yes
        \item No
    \end{enumerate}
    \questionitem Your age: \underline{\hspace{2cm}} years old.
    \questionitem Name of your high school: \underline{\hspace{4cm}}.
\end{enumerate}

\section*{Part 2: Application Preferences and Admission Results}

\begin{enumerate}[leftmargin=*]
    \questionitem Do you have a most preferred college and major?
    \begin{enumerate}[label=(\alph*)]
        \item Yes
        \item No
    \end{enumerate}
    
    \questionitem If you answered ``Yes'' to Q7, please list your preferred colleges and majors in descending order of preference. For each college, specify up to six majors and whether you would accept adjustment.
    
  \begin{table}[h]
    \centering
    \begin{tabular}{|l|c|c|c|c|c|c|c|}
    \hline
    College name & Major 1 & Major 2 & Major 3 & Major 4 & Major 5 & Major 6 & Accept major change? \\
    \hline
    1st choice & & & & & & & \\
    \hline
    2nd choice  & & & & & & & \\
    \hline
    3rd choice & & & & & & & \\
     \hline
    4th choice & & & & & & & \\
    \hline
    5th choice  & & & & & & & \\
    \hline
    6th choice  & & & & & & & \\
    \hline
    7th choice  & & & & & & & \\
    \hline
    8th choice  & & & & & & & \\
    \hline
    9th choice & & & & & & & \\
    \hline
    10th choice & & & & & & & \\
    \hline
  
    \end{tabular}
    \end{table}
    
    Are there more than 10 colleges that you would like to choose? 
    \begin{enumerate}[label=(\alph*)]
        \item Yes
        \item No
    \end{enumerate}
    
    \questionitem Was the  list of choices you provided in Q8 drawn up before the day of online application, or did you add new colleges/majors on the day of application?
    \begin{enumerate}[label=(\alph*)]
        \item The list was drawn up before the online application day.
        \item I added new colleges/majors on the application day. 
        
        The reason was: \underline{\hspace{4cm}}.
    \end{enumerate}
    
    \questionitem Which college and major were you eventually admitted to? 
    \begin{center}
    College: \underline{\hspace{4cm}} \quad Major: \underline{\hspace{4cm}}
    \end{center}
    
    \questionitem Are you satisfied with your final admission result?
    \begin{enumerate}[label=(\alph*)]
        \item Satisfied
        \item Not satisfied
    \end{enumerate}
\end{enumerate}

\section*{Part 3: Application Strategies}

Please answer the following questions about your application strategy.

\begin{enumerate}[leftmargin=*]
    \questionitem Did you have a strategy for submitting your application?
    \begin{enumerate}[label=(\alph*)]
        \item Yes
        \item No
    \end{enumerate}
    
    \questionitem If yes, what was your strategy? (Multiple answers allowed.)
    \begin{enumerate}[label=(\alph*)]
        \item Start with the most preferred college; whenever my rank falls outside the college's admission quota, move to the next choice, and proceed down the list in descending order until I can be admitted.
        \item Select a college before 11:00, then before the application deadline, choose the best college my score can reach.
        \item First apply to various colleges to learn information about my rank, then before the application deadline, choose the best college my score can reach.
        \item No particular strategy; I chose any college to which I could be admitted.
        \item Other strategy (please describe): \underline{\hspace{4cm}}.
    \end{enumerate}
    
    \questionitem From whom did you mainly receive advice on \emph{how} to apply (i.e., application strategy advice, not recommendations on which colleges/majors to choose)?
    \begin{enumerate}[label=(\alph*)]
        \item School and teachers
        \item Professional agencies
        \item Students and parents who applied in previous years
        \item Online sources (e.g., admission websites, Weibo, WeChat, Douyin, etc.)
        \item I did not receive any advice
        \item Other (please describe): \underline{\hspace{4cm}}.
    \end{enumerate}
\end{enumerate}

\section*{Part 4: Application Behaviors}

The following questions concern your actual application behaviors.
\medskip 

\noindent \emph{Please note: The following questions ask about revisions of your college choice, not revisions of majors. Any change of college in the system counts as one modification (e.g., switching from College A to College B and then back to College A counts as two modifications).}

\begin{enumerate}[leftmargin=*]
    \questionitem How many times did you revise your preferred college?
    \begin{enumerate}[label=(\alph*)]
        \item 0--5 times
        \item 5--20 times
        \item 20--50 times
        \item 50--100 times
        \item More than 100 times
    \end{enumerate}
    
    \questionitem What was the frequency of your revisions? 
    \begin{enumerate}[label=(\alph*)]
        \item Similar frequency throughout the whole window
        \item Most revisions occurred at the beginning of the application window
        \item Most revisions occurred just before the deadline
        \item Most revisions occurred at the beginning and just before the deadline, fewer in between
    \end{enumerate}
    
    \questionitem When did you usually revise your preferred college? (Multiple answers allowed.)
    \begin{enumerate}[label=(\alph*)]
        \item When my rank fell below the admission quota of the college
        \item After hourly release of admission information
        \item After the application deadline of the preceding score band, and before the deadline for my own score band
        \item When others suggested revising
    \end{enumerate}
    
    \questionitem What were the reasons for revising your preferred college? (Multiple answers allowed.)
    \begin{enumerate}[label=(\alph*)]
        \item To obtain current information on the score ranking at a target college
        \item My rank fell below the college's admission quota, so I had to choose another
        \item I found a better choice becoming available
        \item Other (please describe): \underline{\hspace{4cm}}.
    \end{enumerate}
    
    \questionitem How many times did you revise your application in the \textbf{last 1 hour} before the deadline? \underline{\hspace{2cm}} times.
    
    \questionitem How many times did you revise your application in the \textbf{last 10 minutes} before the deadline? \underline{\hspace{2cm}} times.
    
    \questionitem How many minutes before the deadline did you make your \textbf{last} revision? \underline{\hspace{2cm}} minutes.
    
    \questionitem Which of the following best describes your application behavior just before the deadline?
    
    \begin{enumerate}[label=(\alph*)]
        \item I settled early on the best college I could reach and made no further revisions.
        \item I kept refreshing the admission status of my target colleges and switched at the last moment to the most preferred college that my score could reach.
        \item Worried that my rank might fall outside the admission quota at the last moment, I chose a safer college where admission was more secure.
        \item None of the above (please describe): \underline{\hspace{4cm}}.
    \end{enumerate}
    
    \questionitem What information did you consider important when submitting your application? (Multiple answers allowed.)
    \begin{enumerate}[label=(\alph*)]
        \item Information on colleges' admission quotas
        \item My real-time rank at the college/major, and the number of students with the same score
        \item The hourly release of colleges' admission information 
        \item Other (please describe): \underline{\hspace{4cm}}.
    \end{enumerate}
    
    \questionitem Which admission batches did you plan to participate in?
    \begin{enumerate}[label=(\alph*)]
        \item Tier-1 Undergraduate Batch
        \item Tier-1 Undergraduate Batch B
        \item Tier-2 Undergraduate Batch
    \end{enumerate}
\end{enumerate}

\section*{Part 5: Admission System}

\begin{enumerate}[leftmargin=*]
    \questionitem What is your view on the current system of staggered deadlines by score segments? (Single answer.)
    \begin{enumerate}[label=(\alph*)]
        \item It greatly reduces the possibility of being displaced from the selected college at the last moment.
        \item It still leaves the possibility of being displaced from the selected college at the last moment.
        \item It makes it very likely to be displaced from the selected college at the last moment.
    \end{enumerate}
    
    \questionitem Do you think the current application window is long enough?
    \begin{enumerate}[label=(\alph*)]
        \item Yes, because: \underline{\hspace{4cm}}
        \item No, because: \underline{\hspace{4cm}}
    \end{enumerate}
    
    \questionitem If the application deadline were extended by 10 minutes, do you think the admission outcomes would change?
    \begin{enumerate}[label=(\alph*)]
        \item Yes, because: \underline{\hspace{4cm}}
        \item No, because: \underline{\hspace{4cm}}
    \end{enumerate}
    
    \questionitem In your view, are there any problems with the current application system? 
    \vspace{0.8cm}
    \begin{center}
    \underline{\hspace{0.8\textwidth}}\\
    \underline{\hspace{0.8\textwidth}}\\
    \underline{\hspace{0.8\textwidth}}
    \end{center}
\end{enumerate}

\section*{Part 6: Other Information}

\begin{enumerate}[leftmargin=*]
    \questionitem Father's education level:
    \begin{enumerate}[label=(\alph*)]
        \item University or above
        \item Junior college
        \item Technical secondary school, vocational high school
        \item Senior high school
        \item Junior high school
        \item More than 4 years of primary school
        \item 1-3 years of primary school
        \item Illiterate or semi-literate
    \end{enumerate}
    
    \questionitem Father's employer type:
    \begin{enumerate}[label=(\alph*)]
        \item Enterprise
        \item Government agency
        \item Public institution
        \item Other
    \end{enumerate}
    
    \questionitem Father's occupation:
    \begin{enumerate}[label=(\alph*)]
        \item Private entrepreneur
        \item Self-employed
        \item Professional/technical personnel
        \item Head of government or enterprise unit
        \item Department head in government or enterprise
        \item Clerical staff
        \item Skilled worker
        \item Unskilled worker
        \item Commercial and service worker
        \item Farmer
        \item Other worker not classifiable
        \item Homemaker or other non-laborer
    \end{enumerate}
    
    \questionitem Mother's education level:
    \begin{enumerate}[label=(\alph*)]
        \item University or above
        \item Junior college
        \item Technical secondary school, vocational high school
        \item Senior high school
        \item Junior high school
        \item More than 4 years of primary school
        \item 1-3 years of primary school
        \item Illiterate or semi-literate
    \end{enumerate}
    
    \questionitem Mother's employer type:
    \begin{enumerate}[label=(\alph*)]
        \item Enterprise
        \item Party/government agency
        \item Public institution
        \item Other
    \end{enumerate}
    
    \questionitem Mother's occupation:
    \begin{enumerate}[label=(\alph*)]
        \item Private entrepreneur
        \item Self-employed
        \item Professional/technical personnel
        \item Head of government or enterprise unit
        \item Department head in government or enterprise
        \item Clerical staff
        \item Skilled worker
        \item Unskilled worker
        \item Commercial and service worker
        \item Farmer
        \item Other worker not classifiable
        \item Homemaker or other non-laborer
    \end{enumerate}
\end{enumerate}

\end{document}